\documentclass[aps,pre,twocolumn,superscriptaddress,a4paper,showpacs,amsmath,amssymb,floatfix]{revtex4}
\usepackage{graphicx}

\begin{document}

\title{Dynamics of bulk fluctuations in a lamellar phase
studied by coherent X-ray scattering}

\author{Doru Constantin}
\email{constantin@lps.u-psud.fr} \altaffiliation{Permanent
address: Laboratoire de Physique des Solides, Universit\'{e}
Paris-Sud, B\^{a}t. 510, 91405 Orsay Cedex, France.}
\affiliation{\'Ecole Normale Sup\'erieure de Lyon, Laboratoire de
Physique, 46 All\'ee d'Italie, 69364 Lyon Cedex 07, France}

\author{Guillaume Brotons}
\affiliation{Laboratoire
de Physique de l'\'{E}tat Condens\'{e}, Universit\'{e} du Maine,
Facult\'{e} des Sciences et techniques, Av. O. Messiaen -- 72085
Le Mans Cedex 9, France}

\author{Tim Salditt}
\affiliation{Institut f\"ur
R\"ontgenphysik, Georg-August-Universit\"at G\"ottingen,
Friedrich-Hund-Platz 1, D-37077 G\"ottingen, Germany}

\author{\'{E}ric Freyssingeas}
\affiliation{\'Ecole Normale
Sup\'erieure de Lyon, Laboratoire de Physique, 46 All\'ee
d'Italie, 69364 Lyon Cedex 07, France}

\author{Anders Madsen}
\affiliation{European Synchrotron
Radiation Facility, Bo\^{\i}te Postale 220, 38043 Grenoble,
France}

\date{\today}

\begin{abstract}
Using X-ray photon correlation spectroscopy, we studied the layer
fluctuations in the lamellar phase of an ionic lyotropic system. We
measured the relaxation rate of in-plane (undulation) fluctuations
as a function of the wave vector. Static and dynamic results
obtained during the same experiment were combined to yield the
values of both elastic constants of the lamellar phase (compression
and bending moduli) as well as that of the sliding viscosity. The
results are in very good agreement with dynamic light scattering
data, validating the use of the technique in ordered phases.

\end{abstract}

\pacs{61.30.St, 87.15.Ya, 61.10.-i}


\maketitle

\section{\label{Intro}Introduction}

X-ray Photon Correlation Spectroscopy (XPCS) is a relatively new
technique \cite{Gruebel04}, successfully used to study the
dynamics of soft-matter systems, such as colloidal dispersions
\cite{Lal01,Robert05}, fluid interfaces \cite{Kim03,Madsen04} and
free-standing smectic films
\cite{Price99,Sikharulidze02,Sikharulidze03}. Although
conceptually very similar to the traditional Dynamic Light
Scattering (DLS) technique, its main advantages with respect to
DLS are the potential of reaching much higher scattering
wavevectors and the fact that it is much less affected by multiple
scattering.

Among the cited systems, smectic phases are especially adapted to
the use of XPCS techniques, since their high degree of order
confines the scattered signal in the vicinity of the (quasi) Bragg
peaks. To date, these studies were limited to thermotropic smectics
at interfaces: either in thin films \cite{Sikharulidze05}, or at the
interface with air \cite{Madsen03}; in both cases, the dynamics is
driven by the ratio between surface tension and viscosity.

In the present work we use XPCS to measure the dispersion relation
of fluctuations in bulk samples of a lamellar lyotropic phase
(exhibiting smectic symmetry) and compare the results with DLS
measurements. This investigation was mainly prompted by three
questions, which we were able to answer in the affirmative:

\paragraph{Is the technique applicable to these systems?} To our knowledge, XPCS was never applied to lamellar
lyotropic phases; although the symmetry is the same as for
thermotropic smectics, there are notable differences due to the
two-component character of the lyotropic phase (leading to
additional hydrodynamic modes), to its lower elastic moduli, which
influence both the relaxation rates and the `spread' of the
diffuse scattering around the Bragg position (thus limiting the
accessible wave vector range). Finally, the difference in
viscosity and electronic contrast can also have an effect.

\paragraph{Can we determine the intrinsic elastic moduli of the phase} by using bulk
samples? In the smectic systems studied so far, the relaxation was
driven by the surface tension. The sample thickness at which
boundary effects become dominant in the relaxation dynamics
depends on the elastic properties of the phase (more specifically,
on the penetration length). It should be noted that the
compression modulus of lyotropic phases is typically more than
three orders of magnitude below that of thermotropic phases.
Moreover, the bending modulus can be tuned within certain limits
and --in some specific mixtures-- the lamellar spacing can be
easily varied by more than a factor of ten. Thus, lyotropic
smectics provide a much more flexible model system than their
thermotropic counterparts.

\paragraph{Are the results comparable with those obtained by dynamic
light scattering} in terms of accessible range, accuracy etc.? This
is a crucial question, since the main interest of XPCS is the
possibility of complementing and extending the range of DLS
experiments. Such a comparison was already performed for colloid
suspensions \cite{Riese00,Gruebel00}, but not in ordered phases.
Such a comparison is non-trivial for smectic systems, first of all
because XPCS is performed around a Bragg position, while in DLS one
probes the vicinity of the origin of reciprocal space.

\section{\label{Exper}Materials and Methods}

The SDS/pentanol/H$_{2}$O system was extensively used as a model
lamellar phase \cite{Safinya86,Nallet89,Freyssingeas96}. In this
work, instead of pure water we use as solvent a 40/60 (wt\%)
solution of glycerol/H$_{2}$O, in order to increase its viscosity
and correspondingly reduce the relaxation rates of the fluctuations:
$\eta_{\mathrm{sol}} = 3.65 \, \eta_{\mathrm{H_{2}O}}$
\cite{Linde99}. The sample composition by volume is: 19.3 \% SDS,
29.9 \% pentanol and 50.8 \% glycerol/H$_{2}$O. The samples were
prepared in 100 and 200 $\mu$m thick borosilicate glass capillaries
(VitroCom Inc.) and oriented by thermal cycling between the lamellar
and the isotropic phases, resulting in very good homeotropic
anchoring. All measurements were performed at $21.5
\,^{\circ}\mathrm{C}$.

The experiments were performed at the ID10A undulator beamline at
ESRF (Grenoble, France) using an X-ray energy of 13 keV selected by
a Si (111) single-bounce monochromator, in the uniform filling mode
of the storage ring. The beam was defined by a $10 \mu \rm{m}$
pinhole followed by a guard slit for removal of parasitic
scattering. The scattered signal was detected by a fast avalanche
photodiode (APD) and the output signal was processed online by a
FLEX autocorrelator.

\section{\label{Model}Results and Analysis}

\begin{figure}[htbp]
\includegraphics[width=8.5cm]{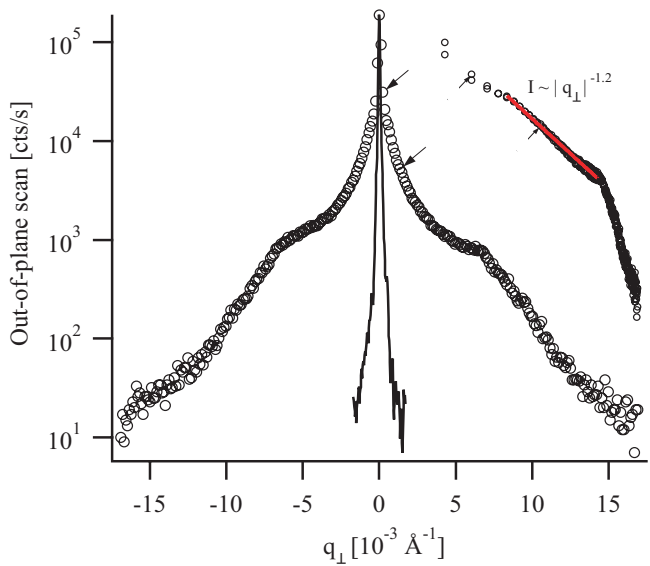}
\caption{(Color online) Out-of-plane scan of the first Bragg peak
(open symbols) shown in log-lin and log-log (inset) representation.
The scan of the primary beam is presented for comparison (solid
line) normalized by a factor $10^{-4}$. In both graphs, the arrows
indicate the range of the XPCS measurements. In the log-log graph,
the straight line is the power-law fit to the intensity (see
text).\label{fig:oop}}
\end{figure}

\subsection{\label{Static}Static scattering}

The smectic periodicity is $d=38.7 \, \mathrm{\AA}$, somewhat
smaller than predicted by the dilution law with pure water, namely
41 {\AA} \cite{Freyssingeas96}. It is not clear if this is due to
the presence of glycerol and how this component influences the phase
diagram of the mixture and the lamellar spacing.

We measured the line shape of the first Bragg peak along the
transverse direction, i.e. in the plane of the layers: $I(q_{\bot})$
(Figure \ref{fig:oop}). More precisely, if we take direction $z$
along the normal to the layers and denote by $x$ the projection of
the incident beam onto the layers (so that the incidence plane is
$(xz)$), the experimental points correspond to taking the detector
out of the plane of incidence, along the $y$ direction: $q_{\bot} =
q_y$.

It is well known that bulk lamellar phases exhibit the
Landau-Peierls instability, leading to a characteristic power-law
variation of the scattered signal close to the Bragg peak
\cite{Caille72} given by $I \sim q_{\bot}^{-(4-2 \eta_{c})}$, with
the conventional Caill\'{e} exponent

\begin{equation}
\label{eq:caille}\eta _{c}= \frac{\pi}{2 d^2} \frac{k_B T}{\sqrt{B
\kappa /d}}
\end{equation}

\noindent where $B$ is the compression modulus of the lamellar phase
\footnote{$B$ stands for the compression modulus at fixed chemical
potential, often denoted by $\overline{B}$ in the literature.} and
$\kappa$ is the bending stiffness of the bilayer.

We observe a clear power-law behaviour out to about $q_{\bot} = 5
\, 10^{-3} \mbox{\AA}^{-1}$, with an exponent of $-1.2$, yielding
$\eta _{c} = 1.4$. This ``coupled elasticity'' regime is followed
by a much steeper decay at higher values, corresponding to length
scales over which the bilayers fluctuate independently; we measure
an exponent of $-3.9$, very close to the theoretical value of
$-4$.

\subsection{\label{Dynamic}Dynamic scattering (XPCS)}

We recorded the time correlation of the diffuse scattered signal in
the vicinity of the first Bragg peak, in the same configuration as
for the static measurements, over a $q_{\bot}$ range indicated by
the arrows in Figure \ref{fig:oop}. When close enough to the peak,
one can separate the scattering vector into two components:
\begin{equation}
\label{eq:sep}
\mathbf{q}=\mathbf{q}_{\mathrm{Bragg}}+\mathbf{q}_{\mathrm{def}}
\end{equation}
with $\left | \mathbf{q}_{\mathrm{def}} \right | \ll \left |
\mathbf{q}_{\mathrm{Bragg}} \right | $, where the Bragg component
shows that the lamellar stack is ``sampled'' with a periodicity
corresponding to the lamellar spacing, and the
$\mathbf{q}_{\mathrm{def}}= \mathbf{q}_{\bot} + q_z
\mathbf{\hat{z}}$ component indicates large-scale superimposed
deformations. In general, for a given deformation vector two
hydrodynamic modes are coupled with the lamellar order: the second
sound \cite{deGennes93} (which relaxes much too fast to be detected
by our setup) and the baroclinic mode, whereby the system fluctuates
at a fixed chemical potential \cite{Brochard75,Nallet89}. This is
the only mode we shall discuss in the following. As
$\mathbf{q}_{\mathrm{def}}$ becomes perpendicular to the z-axis
($\mathbf{q}_{\mathrm{def}} = \mathbf{q}_{\bot}$), the undulation
limit of the baroclinic mode is reached.

The correlation function $g(t)$ was obtained at each $q_{\bot}$
value by acquiring the signal for 1800 or 3600~s. After
normalization by the autocorrelation of the monitor signal and
removal of an oscillatory component due to the mechanical noise of
the setup, $g(t)$ was fitted with the sum of a stretched
exponential (stretching exponent $\beta \sim 0.5$) representing
the relaxation of the undulation mode and a very slow exponential
(decay time $\tau \sim 10 \rm{s}$) of unknown origin:

\begin{equation}
\label{eq:corr} g(t)=1+\left [ a_1 \exp \left [ -(\Omega
t)^{\beta} \right ] + a_2 \exp (-t/\tau) + a_3 \right ]^2
\end{equation}

\begin{figure}[htbp]
\includegraphics[width=8.5cm]{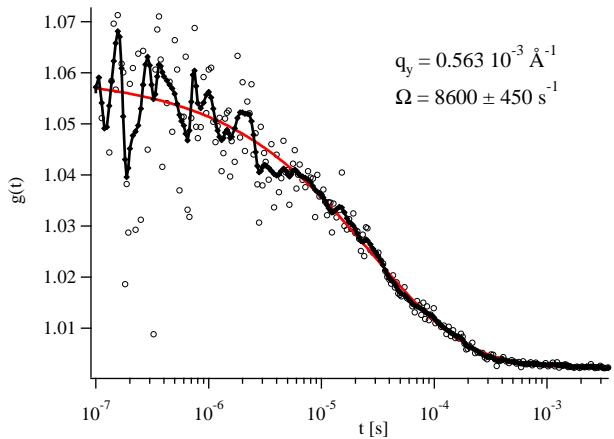}
\caption{(Color online) Correlation function $g(t)$ measured for
($q_{\bot} = 0.563 \, 10^{-3} \mbox{\AA}^{-1}$). Open symbols: raw
signal; dots and solid line: smoothed signal; solid line: fit with
the model described in the text (equation \ref{eq:corr}).
\label{fig:goft}}
\end{figure}

Figure \ref{fig:goft} shows the correlation function $g(t)$
determined for an intermediate value of $q_{\bot}$. Both the raw
signal and the smoothed curve were fitted to the same model; they
yield the same relaxation rate, but the error bars are smaller; in
the following, we only use the smoothed curves.

The dispersion relation $\Omega (\mathbf{q})$ for fluctuations in
the lamellar phase is well known
\cite{Nallet89,Sigaud93,Stepanek01}. In the limit of the undulation
mode, $q_z=0$, it reduces to $\Omega (q_{\bot})=\frac{\kappa /
d}{\eta _3}q^2_{\bot}$, with $\kappa$ the bending stiffness, $d$ the
lattice spacing and $\eta _3$ the layer sliding viscosity
\cite{Martin72}. However, this limit cannot be reached since the
finite size of the capillary (thickness $D = 100\, \mu$m) imposes a
finite $q_z = \pi / D$ component \cite{Ribotta74} that must be taken
into account when describing the dispersion relation, which becomes:

\begin{equation}
\label{eq:disp} \Omega (q_{\bot})=\frac{\kappa / d}{\eta_3}
q^2_{\bot} \left [ 1 + \left ( \frac{\pi}{\lambda D} \right )^2
q^{-4}_{\bot}\right ]
\end{equation}

where $\lambda = \sqrt{\kappa /(dB)}$ is the penetration length of
the smectic phase.

\begin{figure}[htbp]
\includegraphics[width=8.5cm]{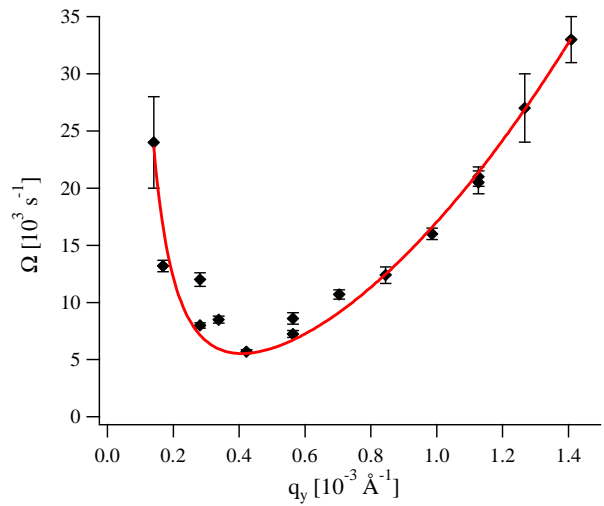}
\caption{(Color online) Measured relaxation rates (diamonds) and fit
with the dispersion relation (Eq. \ref{eq:disp}) shown as solid
line. \label{fig:disp}}
\end{figure}

The fit quality is very good, yielding pa\-ram\-e\-ters $\kappa / (d
\eta_3) = (1.66 \pm 0.06) \, 10^{-10} \, \mbox{m}^{2}/\mbox{s}$ and
$\lambda = (19.6 \pm 0.4) \, \mbox{\AA}$. Using the value of $\eta
_{c} = 1.4$ from the power-law dependence of the static scattering,
we can determine the elastic moduli and the sliding viscosity as:

\begin{equation}
\label{eq:resdyn}
\begin{array}{ccl}
\kappa & = & 2.35 \, 10^{-21} \, \mbox{J} \simeq 0.58 \, k_B T\\
 B & = & 1.56 \, 10^{5} \, \mbox{Pa}\\
\eta_3 & = & 3.65 \, 10^{-3} \, \mbox{Pa s}\\
\end{array}
\end{equation}

First of all, we note that the value found for the sliding viscosity
is exactly that of the solvent: $\eta_3 \simeq \eta_{\mathrm{sol}}$,
as expected. The bending modulus $\kappa$ is similar to that
measured from the dilution law $d(\phi)$ \cite{Freyssingeas96},
which decreases abruptly with the increasing water thickness $d_w$
and saturates at about $0.3 \, k_B T$ for $d_w > 20 \,
\mathrm{\AA}$. In our system, $d_w \simeq 18.5 \, \mathrm{\AA}$, and
a more precise comparison is difficult to make. In the following
section we further check our results against those of light
scattering experiments.

\begin{figure}[htbp]
\includegraphics[width=8.5cm]{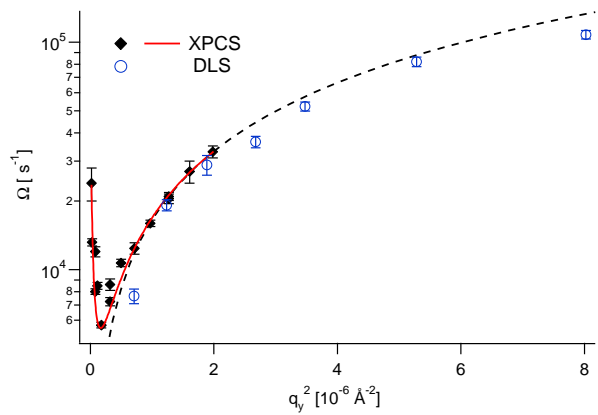}
\caption{(Color online) Comparison between the XPCS and DLS results.
Diamonds and solid line: XPCS data and fit (as in Fig.
\ref{fig:disp}). Dotted line: high-$q_{\bot}$ extrapolation of the
fit. Open dots: DLS results. \label{fig:comp}}
\end{figure}

\subsection{\label{subsec:DLS}Light scattering (DLS)}

The experimental DLS setup is described in reference
\cite{Freyssingeas05}. Briefly, it uses the green (514 nm) emission
line of an Ar laser (Coherent Innova 305) and the scattered signal
is collected with a photon-counting PMT. We only investigated the
undulation mode, corresponding to a scattering vector $q_{\bot}$
contained within the plane of the layers.

The relaxation rate of the undulation mode was measured in DLS for
$q_{\bot}$ between 0.84 and $2.3 \, 10^{-3} \mbox{\AA}^{-1}$; the
experimental points are shown as open dots in Figure \ref{fig:comp}.
They agree very well with the measured XPCS points (solid diamonds)
and the extrapolated dispersion relation for the undulation mode
(dotted line), although they are systematically lower. It is
noteworthy that the correlation functions measured in DLS exhibit
the same stretching exponent $\beta \sim 0.5$ as the XPCS ones;
thus, the stretching is not resolution-induced.

\section{\label{Conc}Conclusion}

We demonstrated the use of the XPCS technique for measuring the
dispersion relation of the undulation mode in a lyotropic lamellar
phase up to a wave vector $q_{\mathrm{max}} = 1.4 \, 10^{-3}
\mbox{\AA}^{-1}$; the results are in very good agreement with
dynamic light scattering measurements. Combining XPCS and static
diffuse scattering measured on the same sample using the same
setup we obtain precise results for the material parameters of the
lamellar phase (Eq. \ref{eq:resdyn}).

In this work, the accessible $q$-range is about half that of
dynamic light scattering. However, in the case of more contrasted
systems with slower dynamics, the DLS range can probably be
exceeded. On the other hand, DLS measurements are also difficult
at {\em low} $q$ values, due to impurities and other defects which
cannot be easily eliminated in ordered systems. The advantage of
XPCS is its selectivity, due to the Bragg ``sampling'' expressed
by Eq. \ref{eq:sep}, which renders it insensitive to such defects.
Thus, it can also be applied to slightly misaligned or ``dirty''
samples.

A systematic comparison between XPCS (measurements around the
Bragg peak) and DLS (probing the origin of reciprocal space) will
probably require a more detailed theoretical description than our
intuitive explanation, especially as the deformation wave vector
$\mathbf{q}_{\mathrm{def}}$ approaches the Bragg value.

The XPCS technique should be particularly interesting for the
study of recently discovered systems, such as DNA-lipid complexes
\cite{Raedler97}, where the dynamics of the confined DNA strands
could provide further insight into the structure of the 2D
`sliding' phase \cite{Salditt97} they form within the host
lamellar matrix, or of inorganic lamellar phases \cite{Gabriel01},
where the presence of heavier elements increases the X-ray
contrast at the same time it hinders light scattering
measurements.

\begin{acknowledgments}
DC received financial support from the CNRS through an "Associated
researcher" fellowship. We acknowledge fruitful discussions with
J\'{e}r\^{o}me Crassous and Fr\'{e}d\'{e}ric Nallet.
\end{acknowledgments}

\bibliography{XPCS}

\end{document}